\documentclass{ieeeaccess}
\usepackage{cite}
\usepackage{amsmath,amssymb,amsfonts}
\usepackage{algorithmic}
\usepackage{graphicx}
\usepackage{textcomp}
\def\BibTeX{{\rm B\kern-.05em{\sc i\kern-.025em b}\kern-.08em
    T\kern-.1667em\lower.7ex\hbox{E}\kern-.125emX}}

\usepackage{bm}
\usepackage{listings}
\lstdefinelanguage{ANTLR}{
    basicstyle=\ttfamily,
    sensitive=true,
    morecomment=[l]{//},
    morecomment=[n]{/*}{*/},
    morestring=[b]",
    morestring=[b]',
	stringstyle=\rmfamily\color[RGB]{128,0,0},
    morekeywords={grammar, import, fragment, returns, locals, throws, catch, finally, mode, options, tokens, channels},
    identifierstyle=\lexerIdStyle
}

\makeatletter
\newcommand*\lexerIdStyle{%
        \expandafter\id@style\the\lst@token\relax
}
\def\id@style#1#2\relax{%
        \ifcat#1\relax\else
                \ifnum`#1=\uccode`#1%
                        \bfseries
                \fi
        \fi
}
\makeatother

\usepackage{mathpartir}

\newcommand{\etal}[1]{#1~{\it et al.}}
\newcommand{\code}[1]{{\small\tt#1}}

\newcommand{\compose}{\bm{\circ}}

\newcommand{\coroutine}[2]{\left [#1; #2\right]}

\newcommand{\Seq}[1]{{\left\langle#1\right\rangle}}

\newcommand{\IDLexer}{\mathbf{ID}}

\newcommand{\Literal}[1]{\mbox{\color[RGB]{128,0,0}\small\tt#1}}
\newcommand{\tlit}[1]{\textrm{#1}}

\newcommand{\Tuple}[1]{{\left ( #1 \right )}}

\newcommand{\ConcreteTypes}{{\mathbf{K}}}
\newcommand{\Provided}{\textrm{provided}}

\newcommand{\Match}{\mathbf{match}}
\newcommand{\First}{\mathbf{first}}
\newcommand{\Hd}{{\mathbf{h}}}
\newcommand{\Tl}{{\mathbf{t}}}

\newcommand{\Require}{{\mathbf{r}}}
\newcommand{\Add}{{\mathbf{a}}}
\newcommand{\Delete}{{\mathbf{d}}}
\newcommand{\Mapping}{{\mathbf{m}}}

\newcounter{composing}
\newenvironment{composingEquation}{\refstepcounter{composing}\equation}{\tag{CR\thecomposing}\endequation}

\newcounter{typing}
\newenvironment{typingEquation}{\refstepcounter{typing}\equation}{\tag{TR\thetyping}\endequation}

\newcommand{\cocome}{CoCoME}

\begin{document}
\history{Date of publication xxxx 00, 0000, date of current version xxxx 00, 0000.}
\doi{10.1109/ACCESS.2023.0322000}

\title{Typing Requirement Model as Coroutines}
\author{\uppercase{Qiqi Gu}\authorrefmark{1}
and
\uppercase{Wei Ke}\authorrefmark{1}}

\address[1]{Faculty of Applied Sciences, Macao Polytechnic University, Macao, China}

\tfootnote{This work is part of the research project (RP/ESAP-02/2017) funded by Macao Polytechnic University, Macao SAR.}

\markboth
{Qiqi Gu \headeretal: Typing Requirement Model as Coroutines}
{Qiqi Gu \headeretal: Typing Requirement Model as Coroutines}

\corresp{Corresponding author: Qiqi Gu (e-mail: qiqi.gu@mpu.edu.mo).}

\begin{abstract}
Model-Driven Engineering (MDE) is a technique that aims to boost productivity in software development and ensure the safety of critical systems.
Central to MDE is the refinement of high-level requirement models into executable code.
Given that requirement models form the foundation of the entire development process, ensuring their correctness is crucial.
RM2PT is a widely used MDE platform that employs the REModel language for requirement modeling.
REModel contains contract sections and other sections including a UML sequence diagram.
This paper contributes a coroutine-based type system that represents pre- and post-conditions in the contract sections in a requirement model as the receiving and yielding parts of coroutines, respectively.
The type system is capable of composing coroutine types, so that users can view functions as a whole system and check their collective behavior.
By doing so, our type system ensures that the contracts defined in it are executed as outlined in the accompanied sequence diagram.
We assessed our approach using four case studies provided by RM2PT, validating the accuracy of the models.

\end{abstract}

\begin{keywords}
coroutine,
model-driven engineering,
static analysis,
type system.
\end{keywords}

\titlepgskip=-21pt

\maketitle

\section{Introduction}

\PARstart{I}{n} the realm of software development, Model-Driven Engineering (MDE) has emerged as a popular paradigm for safety critical systems~\cite{zalila2014methods}, elevating models to the forefront of the process.
MDE enables developers to manage the complexity of software by working at a higher level of abstraction and offers the promise of automatic code generation.
At the heart of this approach is the refinement of high-level requirement models into design models, and eventually down to executable code---without manual intervention~\cite{abrahao2017user}.
Such an emphasis on requirement models underscores their paramount importance, for a misstep at this initial model can spell errors for the entire project.
Thus, careful requirements modeling, paired with early validation and verification, paves the way for clarity and confidence in the envisioned systems~\cite{hofmann2001requirements,zalila2012leveraging}.
One form of requirements validation is to ensure the requirements are consistent~\cite{atladottir2012comparing}.
We firmly believe that by type checking a requirement model, a high level of consistency can be achieved.

\begin{figure*}
\centering
\includegraphics[width=0.8\linewidth]{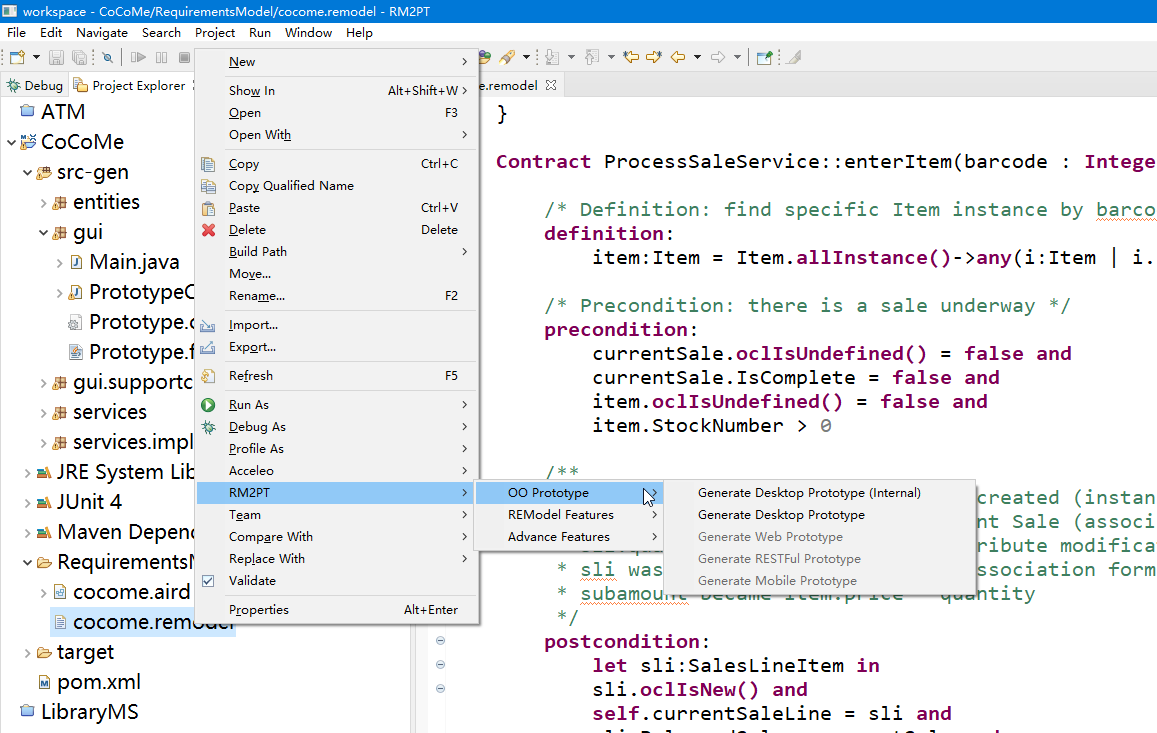}
\caption{In RM2PT, Java source code can be generated from a .remodel file through its context menu}
\label{fig:rm2pt}
\end{figure*}

\etal{Yang}~\cite{yang2019automated} proposed tool RM2PT,
which is a powerful and extensible platform that can generate executable prototypes in the Model-View-Controller (MVC) pattern from requirement models automatically.
The requirement code RM2PT reads is in REModel format.
RM2PT can also visualize requirement code, and the MVC program it generates has GUI for users to view states of the program and check correctness of the model.
Fig.~\ref{fig:rm2pt} exhibits the user interface of the tool.
Users can right-click on a .remodel file, and choose to generate a prototype.
The editor on the right panel is displaying the file \code{cocome.remodel}.

RM2PT is a promising project that inspires a number of research projects.
For example, \etal{Gu}~\cite{gu2022transformation} transforms the output of RM2PT into smart contracts running on Hyperledger Fabric.
\etal{Bao}~\cite{bao2022rm2doc} generates comments for REModel files, and \etal{Yang}~\cite{yang2023deepocl} generates pre- and post-conditions in REModel files from natural languages.
\etal{Reyal}~\cite{reyal2021investigation} provides guidelines of the UI of the generated prototype.

The REModel language is modified from Object Constraint Language (OCL)~\cite{ocl}.
OCL is mainly used for specifying the constraints of UML models including invariants, pre- and post-conditions.
REModel extends OCL in that
a REModel file can define a collection of contracts with such constraints.
Contracts will then be transformed to methods in Java.

REModel includes other useful functional views for software modeling~\cite{ozkaya2020survey}.
For one thing, there is an \code{Interaction} section specifying the order in which the contracts should be called.
This section is rendered as a UML sequence diagram by RM2PT.

Although the RM2PT platform comes with a highlighting service for REModel, the exact syntax of the language is not disclosed,
and there is no type checking on requirement code.
If the output of one contract does not agree with the input type of another contract,
the requirement model is wrong and cannot be used for further development.
Hence, to reduce errors in requirement code and make editing REModel files easier,
this paper aims to add type checking to REModel files, ensuring pre- and post-conditions match the embedded sequence diagram, and help generate integration test cases.
The interplay of pre- and post-conditions in Java methods mirrors the receiving and yielding action of coroutines,
and coroutine is a generalization of function, capable of receiving and yielding data more than once.

\begin{figure*}
\centering
\includegraphics[width=0.7\linewidth]{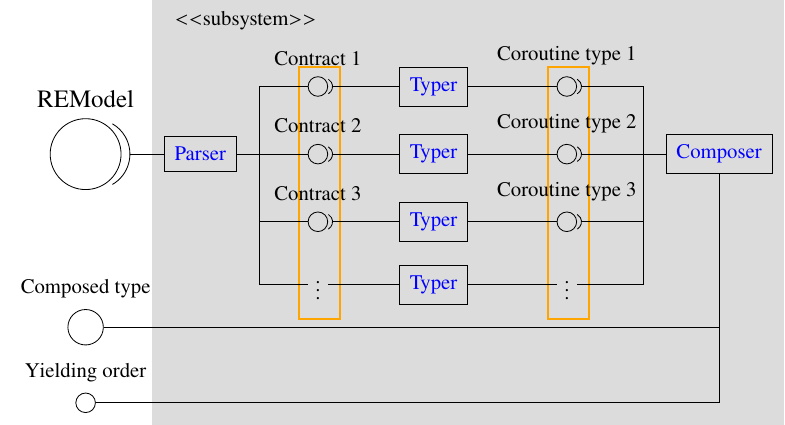}
\caption{The architecture of the requirement model checking framework depicted in the UML component diagram}
\label{fig:architecture}
\end{figure*}

In this paper, we contribute a type system (the gray box in Fig.~\ref{fig:architecture}) for requirement models in MDE.
Our type system creatively types contracts in a requirement model as coroutines by using the Typer component in the figure.
The second contribution is rules for coroutine type composition, i.e., the Composer.
Our rules compose a set of coroutine types into one type to model the collaborative behavior of these coroutines,
and permit users to view a set of coroutines as a single coroutine or function.
The input to our type system is one REModel file, and the output is a composed type and a yielding order, detailed in Fig.~\ref{fig:architecture}.

We firstly formalize the syntax of contracts in REModel with ANTLR 4~\cite{antlr4} lexer and parser rules.
Subsequently we put forward typing rules from the REModel language so we can infer coroutine types from the contracts (bags of pre-conditions and post-conditions) defined therein.
Lastly, we use the composition rules to compose the inferred coroutine types and check whether the contracts are compatible.

The remainder of this paper is organized as follows.
For starters,
Section~\ref{sec:coroutine-type-system} presents composition rules that are invoked in the compose step in Fig.~\ref{fig:architecture}.
Then, Section~\ref{sec:REModel} introduces the syntax of contracts in .remodel files and lists the typing rules, which are the parse step and the type step in the architecture overview.
Section~\ref{sec:applications} demonstrates our coroutine type system, and makes use of the composed type and the yielding order.
Discussion and limitation is in Section~\ref{sec:discussion}.
Related work is reviewed in Section~\ref{sec:related-works}.
Finally we conclude.

\section{Coroutine Type System}
\label{sec:coroutine-type-system}

This paper contributes a novel type system on the REModel language that types contract blocks in this language as coroutine types,
then the coroutine types are composed for verifying the correctness of the requirement model.
This section here presents the rules for type composing because it is more interesting.
These rules can be hooked or listened, and one application is to formulate a potential execution order. This usage is to be discussed in Section~\ref{sec:applications-order}.
Then, Section~\ref{sec:REModel} formalizes the syntax of REModel and lists typing rules.

The syntax of our type expressions is defined in Fig.~\ref{fig:coroutine-syntax}.
A type $t$ can be a concrete type like \tlit{Int} or \tlit{StringBuilder}, a sequence $\Seq{t_1,t_2}$, a tuple $\Tuple{t_1,t_2}$,
or a coroutine $\coroutine{t_1}{t_2}$ where $t_1$ is named the receiving part and $t_2$ the yielding part.
The two parts are a protocol to control when to execute a coroutine.
A coroutine type can take an optional predicate $p$. For instance,
$$
l: \coroutine{\Seq{x,\tlit{BookCopy}}}{\Seq{x,\tlit{BookCopy},\tlit{Reserve}}} \slash x \mathop{<:}  \tlit{User}
$$
reads: ``$l$ is a coroutine which receives $x$ and \tlit{BookCopy}, and yields $x$, \tlit{BookCopy}, and \tlit{Reserve}, where $x$ is a subclass of \tlit{User}.''

\tlit{Int}, \tlit{String}, and the like are regular types we are familiar with, as seen in expressions like $1:\tlit{Int}$ and $\tlit{foo}:\tlit{String}$.
Then, $K$ along with $S,P,R$ should be perceived as a meta-type akin to the kind ($*$) in Haskell~\cite{faxen2002static,chang2017type} or $\tlit{Set}_1$ in Agda~\cite{bove2009brief}.

In abstract syntax,
$p$ stands for one product type, and its (meta-)type is $P$, i.e., $p:P$.
A coroutine type is presented by $\theta$, and its type is $R$, i.e., $\theta:R$.
A sequence type is presented by $\Theta$, and its type is $S$.

A sequence is flat, satisfying the associative law, $\Seq{\Seq{t_1,t_2},t_3} = \Seq{t_1,\Seq{t_2,t_3}}$, where both sides can be simplified to $\Seq{t_1,t_2,t_3}$.
A sequence of a single type $t$ is called a list of $t$, written as $t^n$, where $n$ is the length.
A list of indefinite length is denoted as $t^\ast$, while a list of zero length is denoted as $\varnothing$.
A product functions as a tuple in programming languages.


\begin{figure}
\[
\begin{aligned}
t ::=& 					&\mbox{types} &\ T \\
\mid &\ t \slash p		&\mbox{constrained types} & \\
\mid &\ \Seq{t,t}  		& \mbox{sequence types} &\ S \\
\mid &\ \Tuple{t,t} 		& \mbox{product types} &\ P \\
\mid &\ \coroutine{t}{t} & \mbox{coroutine types} &\ R \\
\mid &\ \tlit{Int} \mid \tlit{String} \mid \cdots & \mbox{concrete types} &\ K \\
\end{aligned}
\]
\caption{Abstract syntax for types in our coroutine type system}
\label{fig:coroutine-syntax}
\end{figure}

Many languages have built-in support for product types. Although product types can be seen as a concrete type from the language,
we baked in native support of tuples because they can define priorities in composition.
In another word, types in a tuple are composed first.
Constrained types make the composition less sensitive to the activation order of coroutines. We will discuss the importance of constraints in depth in Section~\ref{sec:Independence-from-Activation-Order}.

\subsection{Composition Rules}\label{sec:rules}

In this subsection, we introduce the compose function $\compose: S \rightarrow R$ that takes a sequence of types and returns a composed coroutine type.
The input sequence can contain coroutine types and products of coroutines.
Concrete types can only be included in coroutine types $R$.
The return type is $R$ and we do not further reduce it to $K$, so we know the data are computed rather than from a static field.
The compose function $\compose$ consists of rules running in a demand-driven strategy~\cite{papazoglou1984outline}.


A coroutine with empty receiving part and empty yielding part is regarded as useless and should be removed from the argument of $\compose$.
A sequence cannot contain $\varnothing$ either. Composition Rule~\ref{eq:remove-void} lists the reduction regarding removing these identity elements.
$e_1 \Rightarrow e_2$ means an expression $e_1$ evaluates or reduces to another expression or value $e_2$.

\begin{composingEquation}\label{eq:remove-void}
\begin{split}
\compose(\Seq{\Theta_1,\coroutine{\varnothing}{\varnothing},\Theta_2}) &\Rightarrow \compose(\Seq{\Theta_1,\Theta_2}) \\
\compose(\Seq{\Theta_1,\varnothing^\ast,\Theta_2}) &\Rightarrow \compose(\Seq{\Theta_1,\Theta_2}) \\
\compose(\Seq{\Theta_1,\varnothing, \Theta_2}) &\Rightarrow \compose(\Seq{\Theta_1,\Theta_2})
\end{split}
\end{composingEquation}

\ref{eq:compose-tuple} is particularly useful for establishing priorities for coroutines, ensuring that high-priority coroutines---enclosed in a tuple---are composed first.
The reduction applies the function $\mathbf{s}$, which converts a product into a sequence.

\begin{composingEquation}\label{eq:compose-tuple}
\compose\Seq{\Theta_1,p, \Theta_2} \Rightarrow \compose\Seq{\Theta_1,\compose(\mathbf{s}(p)),\Theta_2}
\end{composingEquation}


In a coroutine, the receiving part must be fully satisfied before the yielding part runs.
Interleaving receiving and yielding is impossible.
Hence, variables in the receiving part must be bound for later usage in the yielding part.
We require to run the receiving part first in order to model the worst scenario of a function which needs to receive data before starting to yield.
As a result, yielding to itself will happen no more.

The rest of composition rules need to access or manipulate auxiliary data.
Thus the symbol $\vdash$ is employed. Left to $\vdash$ is the context containing Pending Type $t$ and External Yields $E$.
Right to $\vdash$ is an expression which can be a type or a call to the compose function.
We sometimes use a wildcard item $\cdot$ to indicate an unreferenced item in rules. For example $(\cdot,E)$ means at this point the pending type, which may or may not be $\varnothing$, is not used in other parts of the rule.

\begin{composingEquation}
(\varnothing, E) \vdash \compose(\coroutine{s}{t}) \Rightarrow
(\varnothing, \varnothing) \vdash  \coroutine{s}{\Seq{E,t}}
\label{eq:e-to-one}
\end{composingEquation}

In \ref{eq:e-to-one}, we start with a context that contains an empty Pending Type.
To compose a single coroutine, the yielding part $t$ of the coroutine is prefixed with External Yield $E$, forming a new sequence $\Seq{E,t}$.
The reason for this ordering is that $E$ represents what other coroutines have already yielded.
This rule is a terminal step because on the right-hand side of the $\Rightarrow$ symbol, we don't have another call to $\compose$.
The end result is a single coroutine, matching the definition $\compose: S \rightarrow R$.

The receiving part and the yielding part can be a sequence.
Therefore, we define a head function and a tail function to get the first element and rest elements of a type, shown in Fig.~\ref{head-tail}.
When the yielding part is a sequence, elements are yielded one by one, and the yielded element becomes the pending type.
When the receiving part is a sequence, the head of the sequence is the demand, driving the composition.
For all other types, the head is themselves and the tail is nothing.
To support first-class coroutines,
the head of a coroutine type is itself because we are not willing to break the structure.
If coroutine $a$ yields coroutine $b$, we want to yield $b$ as a whole, rather than only yielding part of $b$.

With the head and tail functions, we are ready to detail the rules with respect to the yield operation and the resume operation.

\begin{figure}
\[
\begin{aligned}
\Hd(k) &= k \qquad & (k \in \ConcreteTypes) \\
\Hd(\coroutine{s}{t}) &= \coroutine{s}{t} \\
\Hd(\Seq{s,t}) &= \Hd(s) \\
\Tl(k) &= \varnothing & (k \in \ConcreteTypes) \\
\Tl(\coroutine{s}{t}) &= \varnothing \\
\Tl(\Seq{s,t}) &= \Seq{\Tl(s),t}
\end{aligned}
\]
\caption{Definition of head $\Hd$ and tail $\Tl$ of a type}\label{head-tail}
\end{figure}

\subsubsection{Yielding}

We use a function $\First$ to find the first coroutine $\theta$ in a list $\Theta$ of coroutines that matches a condition $p$, written as
$(\theta, \Theta_1, \Theta_2) = \First(\Theta, \lambda\theta.p(\theta)),$
where along with the coroutine, $\First$ also returns $\Theta_1$ all elements before $\theta$ and $\Theta_2$ all elements after $\theta$.
If $\First$ cannot find a matching element, $\theta = \Theta_2 = \varnothing$.

\begin{figure*}
\begin{composingEquation}
\begin{split}
(\varnothing, \cdot) \vdash \compose(\Theta) \Rightarrow
  (\Hd(s), \cdot ) \vdash \compose(\Seq{\Theta_1, \coroutine{\varnothing}{\Tl(s)}, \Theta_2})
   \quad\Provided\\
(\theta,\Theta_1, \Theta_2) = \First(\Theta, \lambda \theta. \theta=\coroutine{\varnothing}{s} \wedge s \neq \varnothing)
\end{split}
\label{eq:yield}
\end{composingEquation}
\begin{composingEquation}
\begin{split}
(\varnothing, \cdot )\vdash \compose(\Theta) \Rightarrow
  (\varnothing, \cdot) \vdash \compose(\Seq{\Theta_1, \coroutine{\varnothing}{\Tl(s)}, \coroutine{u}{v}, \Theta_2})
  \quad\Provided \\
(\theta,\Theta_1, \Theta_2) = \First(\Theta, \lambda \theta. \theta=\coroutine{\varnothing}{s} \wedge  \Hd(s)=\coroutine{u}{v})
\end{split}
\label{eq:yield-co}
\end{composingEquation}
\begin{composingEquation}
\begin{split}
(\varnothing, E) \vdash \compose(\Theta) \Rightarrow
  (\varnothing, \varnothing) \vdash \Seq{E,\Theta_1}\quad\Provided\;
  (\varnothing,\Theta_1, \varnothing) = \First(\Theta, \lambda \theta. \theta=\coroutine{\varnothing}{\cdot})
\end{split}
\label{eq:co-to-ext}
\end{composingEquation}
\end{figure*}

\ref{eq:yield} is triggered when there is no Pending Type and
the function $\First$ is able to find the a coroutine whose yielding part is not $\varnothing$.
Then, we transfer the head of its yielding part into the context.
In case a coroutine has exhausted its action statements, its yielding part $\Tl(s)$ would be $\varnothing$ and it's subject to deletion by \ref{eq:remove-void}.

Rather than yielding a concrete type, if the yielded type is a coroutine,
\ref{eq:yield-co} requires to transfer the yielded coroutine into $\Theta$, after where it is from.

\ref{eq:co-to-ext} is used when no coroutine has empty receiving part and Pending Type is $\varnothing$, referred as a deadlock state.
In such cases, the composition result is $E$ followed by all the remaining coroutines from the original list.
$\Theta$ is equal to $\Theta_1$ in this case.

\subsubsection{Resuming}

\begin{figure*}
\begin{composingEquation}
\begin{split}
(t, \cdot) \vdash \compose(\Theta) \Rightarrow
  (\varnothing, \cdot) \vdash \compose(\Seq{\Theta_1, \coroutine{\Tl(s)}{u}[D], \Theta_2 }) \quad \Provided \\
(\theta,\Theta_1, \Theta_2) = \First(\Theta,\lambda \theta. \theta=\coroutine{s}{u} \wedge \Match(t,\Hd(s))=D)
\end{split}
\label{eq:resume}
\end{composingEquation}
\begin{composingEquation}
\begin{split}
(t ,E)\vdash \compose(\Theta)  \Rightarrow (\varnothing, \Seq{E,t}) \vdash \compose(\Theta) \quad\Provided \\
(\varnothing,\Theta_1, \varnothing)=\First(\Theta, \lambda \theta. \theta=\coroutine{s}{\cdot } \wedge \Match(t,s)=\bot)
\end{split}
\label{eq:external}
\end{composingEquation}
\begin{composingEquation}
\begin{split}
(\varnothing, \cdot )\vdash \compose(\Theta)  \Rightarrow (\varnothing, \cdot ) \vdash \compose(\Seq{\Theta_1, \coroutine{\Tl(s)}{u}[D], \Theta_2} -\theta') \quad \Provided\\
(\theta,\Theta_1, \Theta_2) = \First(\Theta,\lambda \theta. \theta=\coroutine{s}{u} \wedge \Hd(s)=\coroutine{\cdot}{\cdot}) \\\textrm{and}\;
(\theta', \cdot, \cdot) = \First(\Theta-\theta,\lambda \theta'. \Match(\theta', \Hd(s))=D)
\end{split}
\label{eq:resume-co}
\end{composingEquation}
\begin{composingEquation}
\begin{split}
(\varnothing, E) \vdash \compose(\Theta) \Rightarrow
  (\varnothing, \Seq{E_1,E_2}) \vdash \compose(\Seq{\Theta_1,\coroutine{\Tl(s)}{u}[D],\Theta_2}) \quad\Provided\\
(\varnothing,\cdot, \cdot) = \First(\Theta, \lambda \theta. \theta=\coroutine{\varnothing}{\cdot}) \;\textrm{and} \\
(t, E_1,E_2)= \First\big(E,\lambda t.
(\theta,\Theta_1,\Theta_2)=\First(\Theta, \lambda \theta. \theta=\coroutine{s}{u} \wedge \Match(t,\Hd(s))=D)\big)
\end{split}
\label{eq:loop-external}
\end{composingEquation}
\end{figure*}

When Pending Type $t$ is not $\varnothing$, the resume operation, \ref{eq:resume}, is triggered.
We find the first coroutine $\theta$ that can receive $t$, and resume it.
$\Match(\cdot,\cdot)$, defined in Fig.~\ref{eq:match}, checks if two types can match and returns conditions $D$.
Conditions $D$ are in the form of variable bindings as $\theta$ may contain variables.
For example $\Match(\tlit{Int}^5,\tlit{Int}^n)=\{n=5\}$. If in no way can two types match, $\bot$ is returned.
The absorbing element $\bot$ joining with any condition is $\bot$.
$\Match$ also handles constraint types by satisfying the constraint $p$.

\begin{figure*}
\begin{align*}
\Match(\coroutine{s}{t}, \coroutine{u}{v} \slash p) &=\begin{cases}
 D & \mbox{if $D=\Match(s,u) \cup \Match(t,v)\;$ and $\;p(D)$}, \\
\bot & \mbox{ otherwise.}
\end{cases} \\
\Match(s, \coroutine{\cdot}{\cdot}) &= \bot \\
\Match(s, t \slash p) &= \begin{cases}
 D & \mbox{if $\exists D.\Hd(t)[D] = s\;$ and $\;p(D)$}, \\
\bot & \mbox{ otherwise.}
\end{cases}
\end{align*}
\caption{Definition of $\Match$ for matching two types.}\label{eq:match}
\end{figure*}

\ref{eq:external} stipulates that if none of the coroutines in $\Theta$ can process the Pending Type $t$, then $t$ gets added to the end of $E$.

\ref{eq:resume-co} outlines how a coroutine receives other coroutines.
Line 2 in \ref{eq:resume-co} finds the first coroutine whose head $\Hd(s)$ is a pattern of coroutine,
then Line 3 aims to match this pattern with another coroutine $\theta'$, returning conditions $D$.
Then we apply the condition to $\coroutine{\Tl(s)}{u}$, and also remove the received coroutine from the result by using the minus sign.
This rule, in combination with \ref{eq:yield-co}, is essential for handling first-class coroutines.

If all coroutines in $\Theta$ cannot yield, \ref{eq:loop-external} loops the types in $E$.
It finds the first type $t$ in $E$ such that one coroutine in $\Theta$ can receive $t$.
This rule is critical in composing contracts in REModel
because developers may require and yield data in mismatching order.
Details are in Section~\ref{sec:typing}.

%
%
%

\section{Analyzing REModel Files}
\label{sec:REModel}

\subsection{Syntax of the REModel Language}

Although Yang claims that REModel files are written in Object Constraint Language (OCL) v2.4~\cite{ocl}, there are many keywords or elements foreign to OCL in the example files~\cite{rm2pt-case-studies}, such as \code{UseCaseModel} and \code{@Description}.

The lexer rules of the REModel Language are almost identical with the standard OCL but REModel uses two successive slashes for comment rather than two dashes~\cite{ocl}.

For parser rules, we start with contract sections. A contract section is similar to the Operation declaration in OCL.
Its abstract structure is crystallized in Fig.~\ref{contract-syntax}.
In terms of the \code{type} element, REModel has a notation to write out enumeration types in
$\tlit{type} ::= \cdots \mid \IDLexer\ \Literal{[}\ \Seq{\IDLexer, \IDLexer, \cdots}\ \Literal{]}$.
One enum type can be found in Listing~\ref{service-definitions}.

\begin{figure}
\begin{equation*}
\begin{split}
& \mathrm{contractDefinition} ::= \\
& \quad	\Literal{Contract}\ \IDLexer\ \Literal{::}\ \IDLexer \\
& \quad	\Literal{(}\ \mathrm{parameterDeclarations}?\ \Literal{)}\\
& \quad		(\Literal{:}\ \mathrm{type})?\\
& \quad		\Literal{\{}\ \mathrm{definitions}?\\
& \quad \quad		\mathrm{precondition}?\\
& \quad	\quad	\mathrm{postcondition}?\ \Literal{\}}
\end{split}
\end{equation*}

\begin{equation*}
\begin{split}
& \mathrm{definitions}::= \\
& \quad	\Literal{definition:}\ \Seq{\mathrm{definition}, \mathrm{definition}, \cdots}
\end{split}
\end{equation*}

\begin{equation*}
\begin{split}
& \mathrm{definition}::= \\
& \quad	\IDLexer\ \Literal{:}\ \mathrm{type}\ \Literal{=}\ \mathrm{factor2Expression}
\end{split}
\end{equation*}

\begin{equation*}
\begin{split}
& \mathrm{precondition}::= \\
& \quad	\Literal{precondition:}\ \mathrm{expression}
\end{split}
\end{equation*}

\begin{equation*}
\begin{split}
& \mathrm{postcondition}::= \\
& \quad	\Literal{postcondition:}\ \mathrm{expression}
\end{split}
\end{equation*}
\caption{The abstract syntax for the contract element}
\label{contract-syntax}
\end{figure}

\begin{figure}
\centering
\includegraphics[width=0.9\linewidth]{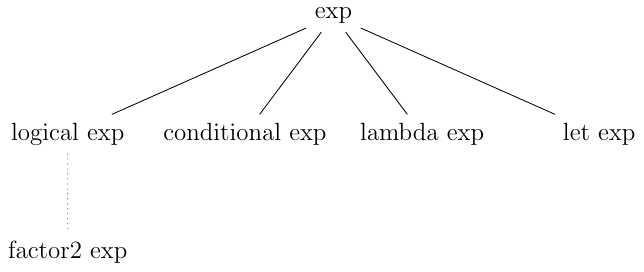}
\caption{Types of expressions in REModel/OCL}
\label{fig:ExpressionsInREModel}
\end{figure}

As Fig.~\ref{fig:ExpressionsInREModel} illustrates, in OCL, an expression element can be either \code{logical\-Expression}, \code{conditional\-Expression}, \code{lambda\-Expression}, or \code{let\-Expression}.
Then, \code{factor2\-Expression} is a component of a \code{logical\-Expression}.

In REModel, \code{let\-Expression} is improved to bear multiple variables.
OCL allows a collection operation, i.e., \code{factor2\-Expression}, to take an iterator, but REModel adopts the Church style~\cite{bove2008dependent} for these lambda expressions, meaning that a type is specified for the iterator, as shown in Fig.~\ref{expression-syntax}.
Moreover, the else part of an if-conditional expression can be left out in REModel.

\begin{figure}

\begin{equation*}
\begin{split}
& \mathrm{letExpression}::=\\
& \quad	\Literal{let}\ \Seq{\IDLexer\ \Literal{:}\ \mathrm{type},  \IDLexer\ \Literal{:}\ \mathrm{type}, \cdots} \\
& \quad	\Literal{in}\ \mathrm{expression}
\end{split}
\end{equation*}

\begin{equation*}
\begin{split}
& \mathrm{factor2Exp}::= \cdots \\
&    \mid \mathrm{factor2Exp}\ \Literal{->any(}\ \mathrm{identifier}\ \Literal{:}\ \mathrm{type}\  \Literal{|}\ \mathrm{expression}\ \Literal{)}\\
&    \mid \mathrm{factor2Exp}\ \Literal{->forAll(}\ \mathrm{identifier}\ \Literal{:}\ \mathrm{type}\ \Literal{|}\ \mathrm{expression}\ \Literal{)}\\
&    \mid \mathrm{factor2Exp}\ \Literal{->select(}\ \mathrm{identifier}\ \Literal{:}\ \mathrm{type}\ \Literal{|}\ \mathrm{expression}\ \Literal{)}
\end{split}
\end{equation*}

\begin{equation*}
\begin{split}
& \mathrm{conditionalExpression}::= \\
& \quad	\Literal{if}\ \mathrm{expression} \\
& \quad \Literal{then}\ \mathrm{expression} \\
& \quad (\Literal{else}\ \mathrm{expression})? \\
& \quad \Literal{endif}
\end{split}
\end{equation*}

\caption{The syntax for expressions in REModel}
\label{expression-syntax}
\end{figure}

Since REModel can compare the initial and the current state of objects, it permits a \code{@pre} tag after an expression, which is defined as
$\mathrm{basicExpression}::= \cdots \mid \mathrm{basicExpression}\ \Literal{.}\ \IDLexer\ \Literal{@pre}?$.

\subsection{Recognizing Identifiers}\label{sec:RecognizingIdentifiers}

Identifiers found in a contract definition can refer to parameters of this contract, fields in this class, and global fields in the system class.
Hence, our type system has to read the service definition sections and extract fields under \code{[TempProperty]}.
A snippet of service definitions is shown in Listing~\ref{service-definitions}.
The service whose name ends with the word System is the system class, fields of which are available for all services.

\begin{lstlisting}[label=service-definitions, float, caption={Two service blocks in REModel}]
Service CoCoMESystem {
	[Operation]
	openCashDesk(cashDeskID)
	closeCashDesk(cashDeskID)
	openStore(storeID)
	closeStore(storeID)

	[TempProperty]
	CurrentCashDesk : CashDesk
	CurrentStore : Store
}

Service ProcessSaleService {
	[Operation]
	makeNewSale()
	enterItem(barcode, quantity)

	[TempProperty]
	CurrentSaleLine : SalesLineItem
	CurrentSale : Sale
	CurrentPaymentMethod : PaymentMethod[CASH|CARD]
}
\end{lstlisting}

Furthermore, we have to know the type inheritance by looking into actor sections.
From Listing~\ref{actor-definitions} we know that the type \tlit{Faculty} is a subtype of the type \tlit{User}.
Given the relationship, if a coroutine type is $\coroutine{\tlit{User}}{\tlit{Book}}$, a type Faculty should be able to activate this coroutine.
The subtype handling is the last step of typing.

\begin{lstlisting}[label=actor-definitions, float, caption={Actor definitions in REModel}]
Actor User {
	@Description( "The user")
	searchBook
	listBookHistory
	makeReservation
	recommendBook
	cancelReservation
}

Actor Faculty extends User {
	@Description( "The faculty user") }
\end{lstlisting}

\subsection{Typing Contracts as Coroutines}
\label{sec:typing}

Equipped with the identifiers and type information of a REModel file, we are able to parse the file and give a coroutine type to each contract block defined therein.
Basically, the precondition element becomes the receiving part of a coroutine, and the postcondition element becomes the yielding part.
We ignore the parameter list and the return type.

Type mapping $\Gamma$ is built according to Section~\ref{sec:RecognizingIdentifiers} which maps identifiers to their types.
However, if an identifier is a global field or a class field, we use the name of the identifier rather than its type in order to distinguish them from data retrieved from other places.

For the contract \code{enterItem} in Listing~\ref{enterItem}, its inferred coroutine type is $[\tlit{CurrentSale},\linebreak[1] \tlit{Item}; \tlit{CurrentSale}, \linebreak[1]\tlit{Item}, \linebreak[1]\tlit{SalesLineItem}, \linebreak[1]\tlit{CurrentSaleLine}]$.
This type is calculated by Typing Rule 1 to 6.

\begin{lstlisting}[label=enterItem, float, caption={The prototype of the enterItem function}]
Contract ProcessSaleService::enterItem(barcode : Integer, quantity : Integer) : Boolean {

	definition:
		item:Item = Item.allInstance()->any(i:Item | i.Barcode = barcode)

	precondition:
		CurrentSale.oclIsUndefined() = false and
		CurrentSale.IsComplete = false and
		item.oclIsUndefined() = false and
		item.StockNumber > 0

	postcondition:
		let sli:SalesLineItem in
		sli.oclIsNew() and
		self.CurrentSaleLine = sli and
		sli.BelongedSale = currentSale and
		CurrentSale.ContainedSalesLine->includes(sli) and
		item.StockNumber = item.StockNumber@pre - quantity and
		sli.Subamount = item.Price * quantity and
		SalesLineItem.allInstance()->includes(sli) and
		result = true
}
\end{lstlisting}


\begin{typingEquation}\label{typing:contract}
\inferrule
  {\Gamma+\Mapping(e_1) \vdash \Require(e_2)=t_1 \\\\ \Gamma+\Mapping(e_1) \vdash \Add(e_3)=t_2 \\\\ \Gamma+\Mapping(e_1) \vdash \Delete(e_3)=t_3 }
  {\Gamma \vdash \parbox{9em}{\Literal{Contract} $\cdots$ \Literal{\{}\\ \Literal{definition:}\;$e_1$ \\\Literal{precondition:}\;$e_2$\\ \Literal{postcondition:}\;$e_3$\Literal{\}}}:\coroutine{t_1}{t_1-t_3+t_2}}
\end{typingEquation}

Typing Rule \ref{typing:contract} stipulates that in order to type a contract block, we need to look into its definition section, precondition and postcondition section.
Function $\Mapping$ extracts type mapping from the definition context.
For listing~\ref{enterItem}, $\Mapping(\code{item:Item=Item.allInstance()...})=\Seq{\tlit{item}: \tlit{Item}}$.
Type \code{Set(t)} in OCL is mapped to $t^\ast$.

The receiving part is typed based on the precondition;
the yielding part is mainly based on the postcondition. What's more, the receiving part has to be added to the yielding part unless the postcondition explicitly removes it.
The plus ($+$) operation stands for sequence joining.
The minus ($-$) operation removes elements from the first operand. Removing an absent element has no effect.
$\Require, \Add, \Delete$ stand for getting required, added, deleted types from given expression. Their definitions can be found in \ref{typing:require}, \ref{typing:add},  \ref{typing:delete}, respectively.

Before we proceed to precondition and postcondition sections, we have to normalize logical expressions into conjunctive normal forms,
and apply \ref{typing:and} to individual terms.
Also let-expressions must be processed by~\ref{typing:let}.

\begin{typingEquation}\label{typing:and}
\openup 2\jot
\begin{split}
\inferrule
  {\Gamma \vdash \Require(e_1)=t_1 \\\\ \Gamma \vdash \Require(e_2)=t_2 }
  {\Gamma \vdash \Require(e_1 \wedge e_2)=t_1+t_2}
\\
\inferrule
  {\Gamma \vdash \Add(e_1)=t_1 \\\\ \Gamma \vdash \Add(e_2)=t_2 }
  {\Gamma \vdash \Add(e_1 \wedge e_2)=t_1+t_2}
\\
\inferrule
  {\Gamma \vdash \Delete(e_1)=t_1 \\\\ \Gamma \vdash \Delete(e_2)=t_2 }
  {\Gamma \vdash \Delete(e_1 \wedge e_2)=t_1+t_2}
\end{split}
\end{typingEquation}

\begin{typingEquation}\label{typing:let}
\openup 2\jot
\begin{split}
\inferrule
{\Gamma+\Mapping \Seq{\IDLexer_1:t_1, \IDLexer_2:t_2, \cdots} \vdash \Require(e)}
{\Gamma \vdash \Require(\Literal{let} \Seq{\IDLexer_1:t_1, \IDLexer_2:t_2, \cdots} \Literal{in}\ e)}
\\
\inferrule
{\Gamma+\Mapping \Seq{\IDLexer_1:t_1, \IDLexer_2:t_2, \cdots} \vdash \Add(e)}
{\Gamma \vdash \Add(\Literal{let} \Seq{\IDLexer_1:t_1, \IDLexer_2:t_2, \cdots} \Literal{in}\ e)}
\\
\inferrule
{\Gamma+\Mapping \Seq{\IDLexer_1:t_1, \IDLexer_2:t_2, \cdots} \vdash \Delete(e)}
{\Gamma \vdash \Delete(\Literal{let} \Seq{\IDLexer_1:t_1, \IDLexer_2:t_2, \cdots} \Literal{in}\ e)}
\end{split}
\end{typingEquation}

For precondition, \ref{typing:require} defines the function $\Require$, standing for ``required''.
Required types translate to the receiving part of the coroutine type.
$\Gamma[e]$ looks up the value of key $e$. For example, $\Seq{\tlit{item}:\tlit{Item}}[\tlit{item}]=\tlit{Item}$.
If operation \code{oclIsUndefined} returns false on an identifier or something must be included in all instances, we know the type of this identifier is required for this contract.
Property checks, such as \code{item.StockNumber > 0} in Listing~\ref{enterItem}, are ignored.

The conjunction and disjunction operator in Boolean algebra are communicative, but our sequence is not.
For instance, the postcondition section in one contract may yield User and Book,
while the precondition section in another may require Book and User, the order reversed.
That's why \ref{eq:loop-external} is crucial for composing contracts
as it can feed the second element Book of the yielding part to the first element of the receiving part.

\begin{typingEquation}\label{typing:require}
\begin{gathered}
\Gamma \vdash \Require(\neg e\Literal{.oclIsUndefined()})=\Gamma[e] \\
\begin{aligned}
\Gamma &\vdash \Require(\neg e_1\Literal{.allInstance()->excludes(}e_2\Literal{)}) \\
	   &=\Gamma[e_2]
\end{aligned}\\
\Gamma \vdash \Require(\neg e)=\varnothing \\
\begin{aligned}
\Gamma &\vdash \Require(e_1\Literal{.allInstance()->includes(}e_2\Literal{)})\\
       &=\Gamma[e_2]
\end{aligned}
\end{gathered}
\end{typingEquation}

\ref{typing:add} is called by \ref{typing:contract} to process the added types for the postcondition section.
It checks if a contract modifies a class property or creates an entity instance.
In detail,
when an object calls \code{oclIsNew()}, this object is for sure newly created and will be added to an entity manager, such as a database or the block on a blockchain.
\code{oclIsNew()} is an OCL operation designed specifically for the post-condition section~\cite{ocl}.
When a class field is assigned, we return the name of the field.
The \code{includes()} and \code{excludes()} operations are handled the same as \ref{typing:require}.

\begin{typingEquation}\label{typing:add}
\begin{gathered}
\Gamma \vdash \Add(e.\Literal{oclIsNew()})=\Gamma[e] \\
\Gamma \vdash \Add((\Literal{self.})?\IDLexer=e_2)=\Gamma[\IDLexer] \\
\begin{aligned}
\Gamma &\vdash \Add(\neg e_1\Literal{.allInstance()->excludes(}e_2 \Literal{)}) \\
		&=\Gamma[e_2]
\end{aligned}\\
\Gamma \vdash \Add(\neg e)=\varnothing \\
\begin{aligned}
\Gamma &\vdash \Add(e_1 \Literal{.allInstance()->includes(} e_2 \Literal{)})\\
 		&=\Gamma[e_2]
\end{aligned}
\end{gathered}
\end{typingEquation}

The postcondition section can contain assertions that an entity must be deleted. These terms in the conjunctive normal form will be matched by \ref{typing:delete}.
As a side note, based on OCL, the includes and excludes operation can take an expression,
but $R_{17}$ and $R_{18}$ in \cite{yang2019rm2pt} put solely an identifier as the argument of the two operations.
Consequently, we assume expressions are not allowed at this position.

\begin{typingEquation}\label{typing:delete}
\begin{gathered}
\Gamma \vdash \Delete(e\Literal{.oclIsUndefined()})=\Gamma[e] \\
\begin{aligned}
\Gamma &\vdash \Delete(e_1\Literal{.allInstance()->excludes(} e_2 \Literal{)})\\
		&=\Gamma[e_2]
\end{aligned}
\end{gathered}
\end{typingEquation}

The coroutine type returned by \ref{typing:contract} is not final yet.
If any type in the receiving part is a super type, we replace all occurrences of this type in the coroutine type with a variable, and add an upper type bound constraint $<:$~\cite{layka2015scala}.
Retrieving child-parent relationship has been covered in Section~\ref{sec:RecognizingIdentifiers}.

\section{Applications}
\label{sec:applications}

We created the concrete syntax of REModel by extending the syntax of OCL retrieved from ANTLR's GitHub repository\footnote{https://github.com/antlr/grammars-v4/tree/master/ocl}.
Our coroutine type system was developed in .NET Core 3.1 (C\#).
The source code, .remodel files for testing, and test cases are all available on GitHub\footnote{https://github.com/gqqnbig/coroutine-program/}.
The \code{RequirementAnalysis} folder implements the REModel parser and typing rules (typer), and the \code{GeneratorCalculation} folder implements composition rules (composer).

Our first experiment is to type four case studies provided by~\cite{yang2019rm2pt} to demonstrate the validity and capability of our system.
The Composer component is indeed standalone. Hence, the second experiment is using Composer to model other programming languages.
This shows the versatility of our type system.

\subsection{Using Yielding Order}
\label{sec:applications-order}

The RM2PT repository~\cite{rm2pt-case-studies} provides a number of case studies.
The case studies have been used in \cite{yang2019rm2pt} to validate the RM2PT platform.
Since our type system is used with the platform, it is natural to employ the same case studies to validate our invention.
The used case studies are Supermarket System (CoCoME), Library Management System (LibraryMS), Automated Teller Machine (ATM), and Loan Processing System (LoanPS);
each has a .remodel file.

We type the four requirement models as coroutines and check whether the inferred coroutine types are correct.
Listing~\ref{cocome-output} is a portion of the typing log when we process the {\cocome} model.
Line 1 to 11 show the type of each contract. (Some trivial contracts are deleted to save space.)
Line 13 is the composed type from the types above, by our composition rules.
This composed type means after running these contracts, the system will have set CurrentStore, CurrentCashDesk, Sale, and so on.

In the meantime, since the yielding order during composition tells the time when the prerequisite of a coroutine has met and the coroutine is ready to execute,
we hooked the execution of \ref{eq:remove-void} and \ref{eq:yield}.
Line 15 to 26 are the execution order of the aforementioned coroutines;
namely we start with createStore, then compose with openStore, and all the way to deleteItem.
Some contracts appear a couple of times because each yielding is recorded and we usually only need to look at the first occurrence.

The execution order is useful in that firstly, for requirement verification, users can check it against the sequence diagram in a requirement model.
If the two orders do not match, this requirement model is contradictory.
Secondly, users can use the order to develop integration tests.
When we were working on paper~\cite{gu2022transformation}, we did not have this coroutine type system and had to manually figured out the calling order in their integration tests if the sequence diagram was not complete.

\begin{lstlisting}[label=cocome-output, float=t, numbers=left, xleftmargin=2em, caption={Partial typing log of the {\cocome} requirement model}]
openStore: [Store; <Store, CurrentStore>]
openCashDesk: [<CashDesk, CurrentStore>; <CashDesk, CurrentStore, CurrentCashDesk>]
makeNewSale: [CurrentCashDesk; <CurrentCashDesk, Sale, CurrentSale>]
enterItem: [<CurrentSale, Item>; <CurrentSale, Item, SalesLineItem, CurrentSaleLine>]
makeCashPayment: [CurrentSale; CashPayment]
createStore: [Void; Store]
deleteStore: [Store; Void]
createCashDesk: [Void; CashDesk]
deleteCashDesk: [CashDesk; Void]
createItem: [Void; Item]
deleteItem: [Item; Void]

[Void; <CurrentStore, CurrentCashDesk, Sale, CashPayment, SalesLineItem, CurrentSaleLine>]

createStore -> openStore ->
openStore -> openStore ->
createCashDesk -> openCashDesk ->
createItem -> openCashDesk ->
openCashDesk -> openCashDesk ->
makeNewSale -> makeNewSale ->
makeNewSale -> makeNewSale ->
enterItem -> enterItem ->
makeCashPayment -> makeCashPayment ->
enterItem -> enterItem ->
enterItem -> deleteStore ->
deleteCashDesk -> deleteItem
\end{lstlisting}

To provide a negative test case,
suppose the model author forgot to include \tlit{CurrentSale} in the \code{enterItem} contract in Listing~\ref{enterItem},
the type of \code{enterItem} would be
$\coroutine{\tlit{Item}}{\Seq{\tlit{Item}, \tlit{SalesLineItem}, \tlit{CurrentSaleLine}}}$.
This type would be composed before \code{makeNewSale}, creating a contrast to the sequence diagram.

\subsection{Using the Composer itself}\label{sec:Independence-from-Activation-Order}
Our composition rules are not necessarily bundled with the type system, but can be used by themselves.
In this subsection we demonstrate the use of the composer to solve a Prolog query.
We borrow a family knowledge base snippet found in~\cite{levesque2012thinking} and list the rules in Listing~\ref{prolog}.
Given a query \code{parent(X,jane), male(X).}, Prolog will tell \code{X=sam}.
Unlike Prolog, our composer cannot return a solution, but it is capable to prove whether a given answer is true or false.

\begin{lstlisting}[language=Prolog,float=t, label=prolog, caption={A family knowledge base in Prolog}]
child(john,sue). child(john,sam).
child(jane,sue). child(jane,sam).
child(sue,george). child(sue,gina).

male(john). male(sam). male(george).
female(sue). female(jane). female(june).

parent(Y,X) :- child(X,Y).
father(Y,X) :- child(X,Y), male(Y).

opp_sex(X,Y) :- male(X), female(Y).
opp_sex(Y,X) :- male(X), female(Y).

grand_father(X,Z) :- father(X,Y), parent(Y,Z).
\end{lstlisting}

Apparently the REModel type system does not recognize the Prolog syntax,
so we assume there is a hypothetical type system that gives a coroutine type for each Prolog rule, and the typing result is given in Fig.~\ref{fig:prolog-types}.

\begin{figure}
\begin{align*}
\Theta=\langle \mathit{child1}:& \coroutine{\Tuple{\mathrm{Child}, \mathrm{John}, \mathrm{Sue}}}{\varnothing},\\
\mathit{child2}: & \coroutine{\Tuple{\mathrm{Child}, \mathrm{Jane}, \mathrm{Sue}}}{\varnothing},\\
\mathit{child3}: & \coroutine{\Tuple{\mathrm{Child}, \mathrm{Sue}, \mathrm{George}}}{\varnothing},\\
\mathit{child4}: & \coroutine{\Tuple{\mathrm{Child}, \mathrm{John}, \mathrm{Sam}}}{\varnothing},\\
\mathit{child5}: & \coroutine{\Tuple{\mathrm{Child}, \mathrm{Jane}, \mathrm{Sam}}}{\varnothing},\\
\mathit{child6}: & \coroutine{\Tuple{\mathrm{Child}, \mathrm{Sue}, \mathrm{Gina}}}{\varnothing},\\
\mathit{childOther}: & \coroutine{\Tuple{\mathrm{Child}, x,y}}{\mathrm{No}} / \Tuple{x,y} \notin \\
					& \big\{
					\Tuple{\mathrm{John}, \mathrm{Sue}},
					\Tuple{\mathrm{Jane}, \mathrm{Sue}},
					\Tuple{\mathrm{Sue}, \mathrm{George}},\\
					&\Tuple{\mathrm{John}, \mathrm{Sam}},
					\Tuple{\mathrm{Jane}, \mathrm{Sam}},
					\Tuple{\mathrm{Sue}, \mathrm{Gina}}	\big\},\\
\mathit{male1}: & \coroutine{\Tuple{\mathrm{Male}, \mathrm{John}}}{\varnothing},\\
\mathit{male2}: & \coroutine{\Tuple{\mathrm{Male}, \mathrm{Sam}}}{\varnothing},\\
\mathit{male3}: & \coroutine{\Tuple{\mathrm{Male}, \mathrm{George}}}{\varnothing},\\
\mathit{maleOther}: & \coroutine{\Tuple{\mathrm{Male}, x}}{\mathrm{No}}  \slash x \notin \left\{\tlit{John}, \tlit{Sam}, \tlit{George}\right\},\\
\mathit{parent}: & \coroutine{\Tuple{\mathrm{Parent}, y, x}}{\Tuple{\mathrm{Child}, x, y}},\\
\mathit{query}: & \coroutine{x}{\Seq{\Tuple{\mathrm{Parent}, x, \mathrm{Jane}}, \Tuple{\mathrm{Male}, x}, \mathrm{Yes}}}, \\
\mathit{answer}:& \coroutine{\varnothing}{\mathrm{Sam}} \rangle
\end{align*}
\caption{The coroutine representation of the family knowledge base in Prolog}
\label{fig:prolog-types}
\end{figure}

We have $\compose(\Theta)=\coroutine{\varnothing}{\mathrm{Yes}}$.
On the other hand, if we replace $\mathit{answer}$ to $\coroutine{\varnothing}{\mathrm{Sue}}$,
\ref{eq:co-to-ext} triggers and $\compose(\Theta)$ deadlocks, so we know Sue is not a solution.

One point of consequence is the presence of constrained types in Fig.~\ref{fig:prolog-types}.
They help the composition result independent from the type order in $\Theta$.
Specifically, if we remove the constraint from $\mathit{childOther}$ and it becomes $\coroutine{\Tuple{\mathrm{Child}, x,y}}{\mathrm{No}}$, and it is placed before $\mathit{child1}$,
then $\mathit{childOther}$ will compose with $\mathit{parent}$ right away, yielding No as a result.

\section{Discussion and Limitation}
\label{sec:discussion}



We ran the REModel type system against four case studies of RM2PT.
The process did not throw exceptions and we manually verified the typing results because there is not another automation system as ground truth.
We also added tests in GitHub Actions to maximize the confidence of correctness.
To test the composer, we selected 6 use cases out of 4 case studies where \code{library.remodel} and \code{cocome.remodel} each contributed 2 cases.
Totally we composed 43 out of 115 contracts defined in the 4 case studies.
We did not compose all contracts because a REModel file typically had contract \code{addXX}, \code{modifyXX}, \code{queryXX}, \code{deleteXX}. Composing these CRUD operations is not interesting.

Apart from the integration tests,
there are 26 unit tests on the composition engine. All these tests are verified by GitHub and indicated by a green tick next to each commit.

There are two contracts \code{inputCard} and \code{inputPassword} in \code{atm.remodel} that our type system does not process correctly.
The post-condition in \code{inputCard} is a conditional statement (see Listing~\ref{if-condition}) that sets CardIDValidated to true or false based on the card validation result.
However, we have no typing rules in regard to if-conditions because our coroutine type expressions do not express branching.
Consequently, this contract is typed $\coroutine{\varnothing}{\varnothing}$, so it cannot compose with other contracts.
Contract \code{inputPassword} has a similar issue.

\begin{lstlisting}[label=if-condition, float, caption={Conditional expressions are not typed}]
postcondition:
	if	bc.oclIsUndefined() = false
	then
		self.CardIDValidated = true and
		self.InputCard = bc and
	else
		self.CardIDValidated = false and
	endif
\end{lstlisting}

Our work has some limitation.
As we mentioned, our type system does not handle if-conditions well and we do not have mechanism to refine the condition or
add the dependency from a precondition to a postcondition.

Second, our type system only discriminates objects being null or non-null, and value checks such as \code{item.StockNumber > 0} are ignored.
Since a type system only concerns types, a different system or a dependent type system---where type level functions are pervasive---may be needed to understand fine-grained checks.

Last, our type system does not divide coroutine types into groups.
If a requirement model contains contracts for both {\cocome} and ATM, the type system still composes all types together, and the result may not make sense.
When we were typing RM2PT case studies, we had to manually find clean-up contracts (such as $\mathit{closeStore}$ and $\mathit{deleteItem}$), and set them to be composed last by using tuples.
If we should forget to do so, the composer might compose $\mathit{closeStore}$ right after $\mathit{openStore}$, blocking $\mathit{makeNewSale}$ and subsequent operations.
An upper stream component will be handy to filter and group contracts, before calling the composer.

Coroutines are ideal to present side-effects, i.e., changes to global data or performs IO~\cite{conway1963design,knuth1997art}, and do not capture parameters of a function and the return value.
Therefore we do not have typing rules reading the parameter list of a contract.
Pure functions, such as query operations, are typed to have the identical receiving part and yielding part because pure functions solely require something in memory but do not change it.
To model parameters and return values, we should use the traditional function types $f: a \rightarrow b$.

\section{Related Work}
\label{sec:related-works}

Model-driven engineering (MDE) can improve implementation productivity.
TopCased~\cite{berthomieu2009formal} is a MDE platform providing modeling languages and code generation. It is widely used in Europe.
RM2PT~\cite{yang2019rm2pt} implements model-to-text (M2T) transformation~\cite{brambilla2017model}, and cuts coding time from hours to one second.
By using MetaCase, another piece of MDE software, a sports watch company observed productivity increase of 750\%~\cite{karna2009evaluating}.

A model in MDE simply refers to an abstraction of data or behavior that captures knowledge, enables automation, and facilitates communication~\cite{evans2014domain}.
A file can be a requirement model; the format a requirement model must conform to is also a model, or we say metamodel.
In this sense, languages are metamodels~\cite{paige2014tutorial,paige2016evolving}.
A valid model is said to conform to its metamodel if it has no syntax errors and fulfills other constrains in the metamodel.
For a .remodel file, it must at least pass syntax check of the REModel language, and its pre- and post-conditions match the specification of the Interaction section, so on and so forth.

REModel is basically a variation of OCL.
OCL is typed, with basic types such as Integer, Boolean, and String, and parameterized sets such as Set(T) and Sequence(T), as well as tuples~\cite{ocl}.
The concrete types in our coroutine type system map to the basic types in OCL.
\etal{Distefano}~\cite{distefano2000towards} gives a formal semantics to OCL invariants and pre-, post-conditions, and run model checking of object-oriented programs.
Nevertheless, their approach does not support type inheritance.
Reference~\cite{combemale2008property} allows to write temporal properties in OCL, for instance work 2 will start only when work 1 is finished.
These properties can be checked by Tina model checker~\cite{berthomieu2004tool}.
Reference~\cite{zalila2012leveraging} verifies behavioral properties of a model with Petri Nets and can give a counterexample if a property (assertion) fails.
A formal approach is proposed by \cite{rusu2011embedding} to define and analyze domain-specific modeling languages (DSML). It represents DSML metamodels and their conforming models as a Maude specification~\cite{clavel2007all}.
Fiacre~\cite{berthomieu2008fiacre} is a formal specification language that targets both the behavioral and timing aspects of real-time systems.
Its authors transform UML and OCL to Fiacre, and then perform Tina verification.
Overall, prior work mainly studies constraints in a single OCL context. In contrast, our work focuses on the interplay of a set of OCL contexts (named contract in REModel), and
can work out temporal properties from pre- and post-conditions by modeling contract blocks as coroutines.

Coroutines are included in assorted modern programming languages.
In particular, coroutines in Kotlin can dispatch user actions faster
comparing to Java threads~\cite{chauhan2021performance}.
Vector~\cite{vector-android} is one of the Model-View-Intent libraries that utilize the coroutine backend.
Our coroutine type composition rules are to some extend similar to Vector and other coroutine dispatchers and reducers,
but we reduce types rather than values, and have limited capability in arithmetic and logical calculations.
We do not concern performance either.

%
%

A coroutine system can be categorized into three aspects, namely control-transfer mechanisms, whether coroutines are first-class objects,
and whether the control can be suspended within nested calls~\cite{moura2009revisiting}.
In REModel, when a contract calls a built-in function, the function cannot stop the whole contract; hence REModel is a stackless coroutine language.
Overall, our type composition rules support asymmetric, first-class, stackless coroutines.
Kotlin Coroutines are stackless as well because not every function has access to the CoroutineScope of the parent coroutine.
\etal{Ikebuchi}~\cite{ikebuchi2022certifying} proposes a high-level language for defining coroutines and the language can be compiled to low-level C code.
This approach is useful for implementing security-critical network protocols.
Nevertheless, this paper concentrates on the execution logic of a coroutine rather than an overall type of a coroutine, not mention composing.
The coroutine type in \cite{prokopec2018theory} has three elements, parameters $P$, return type $R$, and yielded values $Y$.
It does not include received values because the resume site creates a new instance every time and pass the received values as parameters.

Our type system can be perceived as an integration platform and G{\'o}rski~\cite{gorski2021one} proposed 6 views for such system, including Integrated Services view, and Contracts view.
Our Fig.~\ref{fig:architecture} is indeed the Integrated services view,
where each rectangle box is G{\'o}rski's contract.
His Contracts view illustrates the cooperation of components, and each contract has a provided interface (a ball) and a required interface (a socket).
G{\'o}rski's contract is dissimilar to the contract keyword in REModel.
In REModel, a contract keyword denotes a function, of which the function body is fully specified by pre-conditions and post-conditions.
A coroutine with void receiving part means it does not receive data; a coroutine with void yielding part means it does not yield data.
Therefore we adopt $\varnothing$, the void type in type theory, in our type system to present a type with no inhabitants~\cite{constable1991type}~\cite{norell2007towards}.
Besides validating requirement models, a type system can check correctness of format string~\cite{weitz2014type}, regular expressions~\cite{spishak2012type}, and so on.

\section{Conclusion}
\label{sec:conclude}

In software engineering, requirements validation is the process of checking whether requirements meet the customers' real needs.
It is critically important because requirements errors will lead to extensive rework if those problems are discovered during the later phases of the software development.
To verify a requirement file, RM2PT implements a transformation from requirement files to executable prototypes so that developers can verify the requirement by running the prototype.
However, it does not have powerful verification tools for models themselves.
To compensate for the shortcoming, this paper takes another approach. Rather than transforming a requirement file,
we directly verify the file by a formal method to ensure consistency.
The file is written in the REModel language, which is a variation of OCL.

We contribute a new type representation and computation method for combining asymmetrical, first-class, stackless coroutines.
By formally defining the syntax of REModel, our type system checks each contract section in REModel and infers a coroutine type based on pre- and post-conditions.
Then a selection of coroutine types can be combined into a single coroutine type to model the final result of a sequence of operations.
The composition operation is expected to produce the same sequence diagram embedded in the requirement model.
If not, discrepancies have been spotted.

We evaluated our approach on the four official case studies of RM2PT, namely ATM, CoCoME, LibraryMS, and LoanPS, and confirmed that the coroutine composition results are correct.
Moreover, with constrained types and the composition rules,
coroutine types can model other programming languages, such as Prolog.
The types are composed in a way independent from the activation order.
As a result, no matter the clause order of a logical expression in pre-condition or post-condition, contracts are composed to the same type.
Subtypes are supported as well.

There are some future work.
Firstly, we plan to revise the system to handle fine-grained checks, such as conditional expressions.
Secondly, an extra upper stream component can do its part to divide coroutines into groups, or assign priorities to coroutines or contract.
Finally, we hope to adapt our coroutine type system for a wider range of requirement models, not only for RM2PT.

\bibliographystyle{IEEEtran}
\bibliography{mybibliography.bib}

\begin{IEEEbiography}[{\includegraphics[width=1in,clip,keepaspectratio]{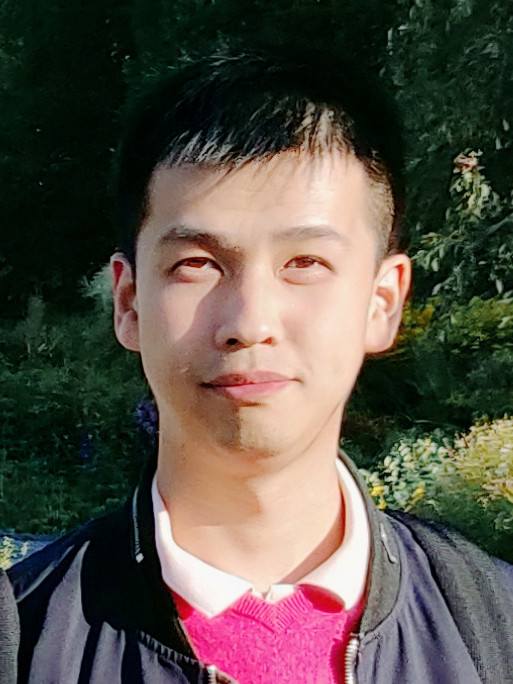}}]{Qiqi Gu}
received his Master of Science degree in computer science from University of California, Los Angeles in 2015.
He worked as software development engineer in Irvine, California.
Currently he is pursuing the PhD degree in computer science at Macao Polytechnic University in Macao, China.
His research interests are software engineering, especially on Model-Driven Engineering and blockchain applications.
\end{IEEEbiography}

\begin{IEEEbiography}[{\includegraphics[width=1in,clip,keepaspectratio]{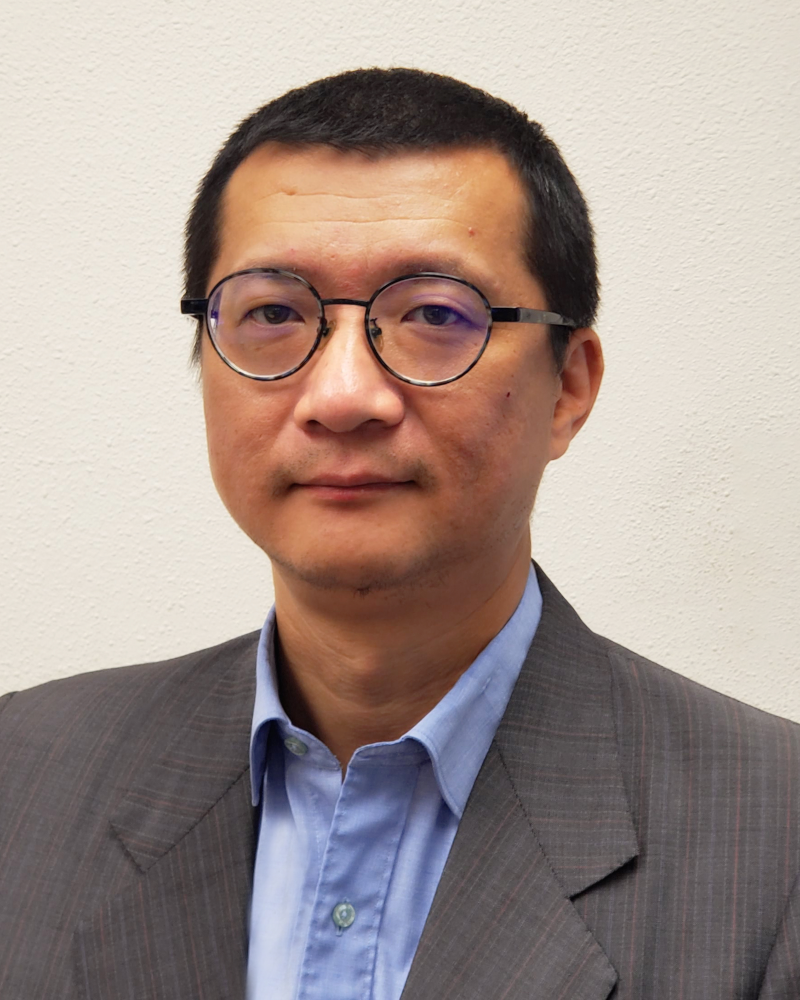}}]{Wei Ke}
received his PhD degree from School of Computer Science and Engineering, Beihang University. He is an associate professor of Computing Program, Macao Polytechnic University. His research interests include programming languages, image processing, computer graphics and component-based engineering and systems. His recent research focuses on the design and implementation of open platforms for applications of computer graphics and pattern recognition.
\end{IEEEbiography}

\EOD

\end{document}